\documentclass[useAMS,usenatbib]{mn2e}

\usepackage{graphicx}
\usepackage[T1]{fontenc}
\usepackage{amssymb}
\usepackage{color}

\newcommand{\url}[1]{\texttt{#1}}

\newcommand{\noun}[1]{\textsc{#1}}
\newcommand{\simapp}{\mathord{\sim}}
\newcommand{\gtrsimapp}{\mathord{\gtrsim}}
\newcommand{\gtrapp}{\mathord{>}}
\newcommand\ion[2]{#1$\;${\small \uppercase\expandafter{\romannumeral #2}}\relax}%
\newcommand\nodata{ ~$\cdots$~ }%

\title[Five New AM CVn Systems from the PTF]{Five New Outbursting AM CVn Systems Discovered by the Palomar Transient Factory}

\author[D. Levitan et al.]{David Levitan$^1$\thanks{E-mail: dlevitan@caltech.edu}, 
Thomas Kupfer$^2$,
Paul J. Groot$^{1,2}$,
Shrinivas R. Kulkarni$^1$,\newauthor
Thomas A. Prince$^1$,
Gregory V. Simonian$^1$,
Iair Arcavi$^3$,
Joshua S. Bloom$^4$,\newauthor
Russ Laher$^5$,
Peter E. Nugent$^{4,6}$,
Eran O. Ofek$^3$,
Branimir Sesar$^1$,
and Jason Surace$^5$.
\\ \\
$^1$ Division of Physics, Mathematics, and Astronomy, California Institute of Technology, Pasadena, CA 91125, USA.\\
$^2$ Department of Astrophysics/IMAPP, Radboud University Nijmegen, PO Box 9010, NL-6500 GL Nijmegen, the Netherlands.\\
$^3$ Department of Particle Physics and Astrophysics, Weizmann Institute of Science, Rehovot, 76100, Israel.\\
$^4$ Department of Astronomy, University of California, Berkeley, CA 94720-3411, USA.\\
$^5$ Spitzer Science Center, MS 314-6, California Institute of Technology, Pasadena, CA 91125, USA.\\
$^6$ Computational Cosmology Center, Lawrence Berkeley National Laboratory, 1 Cyclotron Road, Berkeley, CA 94720, USA.
}

\begin{document}

\date{Accepted ...  Received ...; in original form ...}

\pagerange{\pageref{firstpage}--\pageref{lastpage}} \pubyear{2012}

\maketitle

\label{firstpage}

\begin{abstract}
We present five new outbursting AM CVn systems and one candidate discovered as part of an ongoing search for such systems
using the Palomar Transient Factory (PTF). This is the first large-area, systematic search for AM CVn systems using only large-amplitude photometric
variability to select candidates. Three of the confirmed systems and the candidate system were discovered as part of the PTF transient search.
Two systems were found as part of a search for outbursts through the PTF photometric database. We discuss the observed characteristics of each of
these systems, including the orbital periods of two systems. We also consider the position of these systems, selected in a colour-independent survey,
in colour-colour space and compare to systems selected solely by their colours. We find that the colours of our newly discovered systems do not
differ significantly from those of previously known systems, but significant errors preclude a definitive answer.
\end{abstract}

\begin{keywords}
accretion, accretion discs -- binaries: close -- novae, cataclysmic variables -- stars: individual: (PTF1\,J043517.73+002940.7, PTF1\,J085724.27+072946.7, PTF1\,J094329.59+102957.6, PTF1\,J152310.71+184558.2, PTF1\,J163239.39+351107.3, PTF1\,J221910.09+313523.1) -- white dwarfs
\end{keywords}

\section{Introduction}

AM CVn systems are rare, ultra-compact, semi-detached white dwarf binaries with periods ranging from 5 to
65 minutes. First proposed as a class of binary systems over 40 years ago by \citet{1967AcA....17..255S}, fewer than
ten additional systems were discovered in the following 30 years. The availability of wide-area surveys --
first photometric, then spectroscopic, and now synoptic -- has resulted
in the discovery of over 20 systems in just the last decade, yet their rich and complex phenomenological behaviour and evolutionary history
has limited our understanding of these degenerate post-period minimum binaries. AM CVn systems are believed to be one
possible outcome of double degenerate white dwarf evolution, along with R CrB stars, massive single white dwarfs,
and type Ia supernovae \citep{1984ApJ...277..355W,2001A&A...368..939N}.
They are extremely important as strong low-frequency Galactic gravitational wave sources
\citep[e.g., ][]{2004MNRAS.349..181N,2007MNRAS.382..685R,2012arXiv1201.4613N} and are
the source population of the proposed ``.Ia'' supernovae \citep{2007ApJ...662L..95B}.
However, the lack of an accurate population density has complicated their use for understanding
these phenomena. We refer the reader to \citet{2005ASPC..330...27N} and \citet{2010PASP..122.1133S}
for general reviews.

The first 11 AM CVn systems were serendipitous discoveries. Several were initially of interest
as supernova candidates \citep[e.g., ][]{1998IAUC.6983....1J,2003IAUC.8077....1W}. The availability of the
Sloan Digital Sky Survey (SDSS) and, specifically, its spectroscopic database, led to the first
systematic search for AM CVn systems and yielded  7 new systems based on their distinctive helium emission lines
and lack of hydrogen \citep{2005MNRAS.361..487R,2005AJ....130.2230A,2008AJ....135.2108A}.
However, the SDSS spectroscopic database is not complete and spectroscopic
targets are selected based on complex colour criteria and fiber availability. \citet{2009MNRAS.394..367R}
noticed that the known AM CVn systems are clustered in
a relatively sparse area of the colour-colour space and proposed a spectroscopic survey of all
SDSS sources inside a pre-determined colour cut. Although predictions of up to a total of 50 systems in the SDSS
were made \citep[based on ][]{2007MNRAS.382..685R}, the survey, now wrapping up, has found fewer than 10 additional systems
\citep{2009MNRAS.394..367R,2010ApJ...708..456R,CARTERAMCVNS}.

Although successful, such colour-selected spectroscopic surveys are both resource intensive and inherently biased to
the previously known population. AM CVn systems have similar colours to other blue objects, including white dwarfs
(particularly DB white dwarfs) and QSOs. Expanding the colour selection to a wider box appears to offer
a significant increase of the number of candidates with few gains in the number of discovered AM CVn systems.
A way to further down-select candidates, or completely change the selection criteria, is thus
necessary.

AM CVn systems are thought to have distinctly different phenomenology dependent on their orbital periods.
Believed to be binaries hosting degenerate mass donors that have evolved through a period minimum, their orbital periods increase
as angular momentum is lost to gravitational wave radiation.
The most recently formed systems, with periods below $\simapp20\,$min, are in a constant
state of high mass transfer from the donor to the optically thick accretion disc.
They have been referred to as ``high'' state systems and exhibit properties similar to
dwarf novae cataclysmic variables in outburst, including superhumps and absorption line
spectra \citep[e.g., ][]{2006MNRAS.371.1231R,2007MNRAS.379..176R,2011ApJ...726...92F}.

The oldest systems -- those with orbital periods over $\simapp40\,$min and thus low mass transfer rates --
are characterized by their lack of photometric variability and strong helium
emission lines from the optically thin disc. Spectroscopic surveys are primarily
sensitive to systems with these emission lines.

Between these two extremes are the ``outbursting'' systems, thought to have
orbital periods between roughly 20 and 40 minutes. These systems are characterized by their changes
between a high state and a low state, which results in both a luminosity
change of 3--5 magnitudes and (typically) a change in the spectral features
from absorption lines to emission lines. In each state, they generally take on the
properties of that state, with the significant addition of photometric variability in the low state
for some of the systems. This variability was tied to the orbital period
by \citet{2011ApJ...739...68L}, but any link between the photometric
and spectroscopic variability remains to be confirmed by observations of additional
systems. Outbursts typically last on the order of days to weeks, and are recurrent
on timescales of a few months to over a year \citep{2000MNRAS.315..140K,2012MNRAS.419.2836R}.
Large-area synoptic surveys are most sensitive to these outbursting systems and the
first system from such a survey was reported by \citet{2011ApJ...739...68L} using
the Palomar Transient Factory \citep[PTF; ][]{2009PASP..121.1395L}. 

In this paper, we continue the survey work started in \citet{2011ApJ...739...68L} and
report on the discovery of an additional five AM CVn systems and one faint candidate as part of the PTF AM CVn System
Key Project. This search is the first systematic, large-area colour-independent search for AM CVn systems
that relies solely on their large-amplitude, photometric variability to identify candidates.

We introduce the PTF in Section \ref{sec:detection} and describe our AM CVn system detection strategy and data reduction processes.
We report on our discoveries in Section \ref{sec:systems}, including both photometric measurements of the individual systems and period analysis
based on phase-resolved spectroscopy. In Section \ref{sec:Discussion} we discuss the features of these systems
and compare their colours to the colour selection criterion used in the aforementioned spectroscopic survey.

\section{Source Detection and Analysis Process}
\label{sec:detection}

The PTF\footnote{\url{http://www.astro.caltech.edu/ptf}} uses the Samuel Oschin $48^{\prime\prime}$
Schmidt telescope at the Palomar Observatory to image up to $\simapp2,000\deg^{2}$ of the sky per night to a median depth on dark nights
of $R\simapp20.6$ or $g'\simapp21.3$ \citep{2009PASP..121.1395L,2009PASP..121.1334R}. The cadence
of observations is not uniform and has varied from 90\,s to 5\,d, depending on the observational programme conducted
at the time. 

Two pipelines process PTF data. The ``transient'' pipeline uses difference imaging
for the rapid discovery of transient events. Exposures are automatically reduced and processed
within a few hours of acquisition and candidate events, identified using difference photometry,
are analysed by both automated routines and humans \citep{2009PASP..121.1395L}.
Conversely, the ``photometric'' pipeline is designed for accuracy, not speed. In this pipeline, exposures are
processed after the end of the night using aperture photometry (Laher et al. in prep).
Instrumental magnitudes are calibrated to SDSS and instrumental effects,
airmass, and background are de-trended \citep{2012PASP..124...62O}. Finally, light curves are generated
using relative photometry algorithms (Levitan et al. in prep). From this photometric database, we select sources
for follow-up observations. Most sources selected as part of this search were identified as cataclysmic variables (CVs) and are detailed in
Groot et al. (in prep).

We refer to the AM CVn systems presented in this paper as either ``transient'' discovered or
``photometrically'' discovered. The former are those initially selected as supernova candidates and classified
as part of the PTF supernova search. ``Photometrically'' identified systems were found by scanning through light curves for
outbursts. These systems were selected by searching for outbursts of 2\,mag above the median magnitude that
have a second measurement at least 1\,d later that is $5\sigma$ brighter than the median. 

\subsection{Data Reduction Procedures}

All PTF light curves presented in this paper were processed through the photometric pipeline referred to earlier. 
The relative photometry algorithm used for the data in
this paper (both PTF and targeted observations) is a matrix-based least
squares algorithm. The algorithm was described briefly in \citet{2011ApJ...740...65O} and \citet{2011ApJ...739...68L}.
Further details and the specific application to the PTF data will be in Levitan et al. (in prep).
We note that the systematic uncertainty limit of the PTF relative photometry is approximately
6--8\,mmag (based on bright stars with $14.5<m_R<16$).
Errors of $\simapp0.1\,$mag are achieved at $m\simapp19$ and $\simapp0.2\,$mag at $m\simapp21$.

The spectroscopic data were reduced using either standard \noun{iraf} tasks or using optimal extraction \citep{1986PASP...98..609H}
as implemented in the \noun{Pamela} code \citep{1989PASP..101.1032M} as well
as the \noun{Starlink} packages \noun{kappa}, \noun{figaro}, and \noun{convert}. The spectra acquired
for the phase-resolved spectroscopy were all reduced using the latter. Spectra obtained from the red side of
Keck-I/LRIS were processed with \noun{L.A. Cosmic} \citep{2001PASP..113.1420V} due to the large number of cosmic rays.
Photometric data from the Palomar $200^{\prime\prime}$ (P200) and the Nordic Optical Telescope (NOT) were reduced
using standard bias-subtraction and flat-fielding techniques.
PSF photometry as implemented in either \noun{daophot} or \noun{autophotom} was used for measurements and absolute calibration was done
by comparing with SDSS measurements in the same filter.

\subsection{Period Analysis\label{sec:peranal}}

For two of the systems, we attempted to measure the orbital period by looking for the ``S-wave'' using the method in \citet{1981ApJ...244..269N}.
The short orbital periods of AM CVn systems require short exposures, and, for these faint systems, 5--10\,h of time on a 8--10\,m telescope.
The S-wave is thought to be caused by the orbital motion of the accretion disc bright spot, formed by the impact of  transferred mass hitting the disc
\citep[e.g., ][]{WARNER1995}. For each spectrum obtained, we summed the flux for $1000\,\mathrm{km}~\mathrm{s}^{-1}$ on either side of the strongest lines
and divided the two measurements. These set of ratios were analysed for each system using a Lomb-Scargle periodogram
as implemented in \citet{2011ApJ...733...10R}.

The verification of the orbital period was done in two ways. First, the strongest emission lines were co-added for each individual spectrum
and converted into a trailed, phase-binned spectrum. This is essentially a two-dimensional representation with wavelength/velocity on one axis
and time/phase on the other. In this image, we expect to see a sinusoidal S-wave from the hotspot on top of the disc emission as a result of
Doppler shift. The particular lines used varied for each system.

To further verify the orbital period, we transformed the phase-binned spectrum into a Doppler tomogram \citep{1988MNRAS.235..269M}.
Doppler tomograms are essential in the study of semi-detached systems since they concentrate orbital velocity variations in a single location
on a velocity map. The phase, radial velocity, and intensity data were used to produce an image in $(K_x,K_y)$ velocity space.
$(K_x,K_y)=(0,0)$ is the location of the centre of mass. For semi-detached systems, four components are typically seen:

\begin{enumerate}
\item the accretor. For AM CVn systems, the absorption line features of the accretor itself are masked
by the emission lines from the accretion disc and thus not visible in the Doppler tomogram.
However, a narrow emission line feature referred to as the ``central spike'' is visible on some typically longer-period systems
and is believed to be on or near the accretor \citep[see e.g., GP Com;][]{2003A&A...405..249M}.
The extreme mass ratio of AM CVn systems results in the accretor having a low velocity.
\item the accretion disc will extend from a relatively low velocity corresponding to the Keplerian orbital velocity of the outer edge of the disc
to the much higher Keplerian orbital velocity at the radius of the white dwarf.
\item the donor star will have a lower velocity than the outer edge of the accretion disc. From Keplerian orbital mechanics,
\[
v_\mathrm{don} = \left(\frac{2\pi G}{P_\mathrm{orb}}\right)^{1/3} \sqrt{\frac{M_\mathrm{acc}}{(M_\mathrm{acc}+M_\mathrm{don})^{1/3}}},
\]
where $v_\mathrm{don}$ is the velocity of the donor, $G$ is the gravitational constant, $P_\mathrm{orb}$ is the orbital period of the system,
$M_\mathrm{acc}$ is the mass of the accretor, and $M_\mathrm{don}$ is the mass of the donor. If we set, for example,
$M_\mathrm{acc}=0.85M_\odot $ and $M_\mathrm{don}=0.035M_\odot$, as found for the eclipsing AM CVn system SDSSJ0926+3624 in \citet{2011MNRAS.410.1113C}, we find that $v_\mathrm{don}\approx700\,\mathrm{km}~\mathrm{s}^{-1}$. However, the donor is typically not seen
in AM CVn systems due to its much lower luminosity relative to the accretor and disc. The sole exception thus far is the 5.4\,min
orbital period system HM Cnc \citep{2010ApJ...711L.138R}.
\item the hotspot is expected to be on the inner edge of the accretion disc on the Doppler tomogram, at a relatively constant velocity. 
The location relative to the accretor is dependent on the size of the disc, but, for longer-period systems, is typically on the opposite side
of the centre of mass in $(K_x,K_y)$ space. The identification of a well-defined hotspot in the Doppler tomogram is a requirement to
establish the orbital period \citep[e.g., ][]{2006MNRAS.371.1231R}.
\end{enumerate}

\section{AM CVn Systems}
\label{sec:systems}

\begin{figure*}
\begin{centering}
\includegraphics{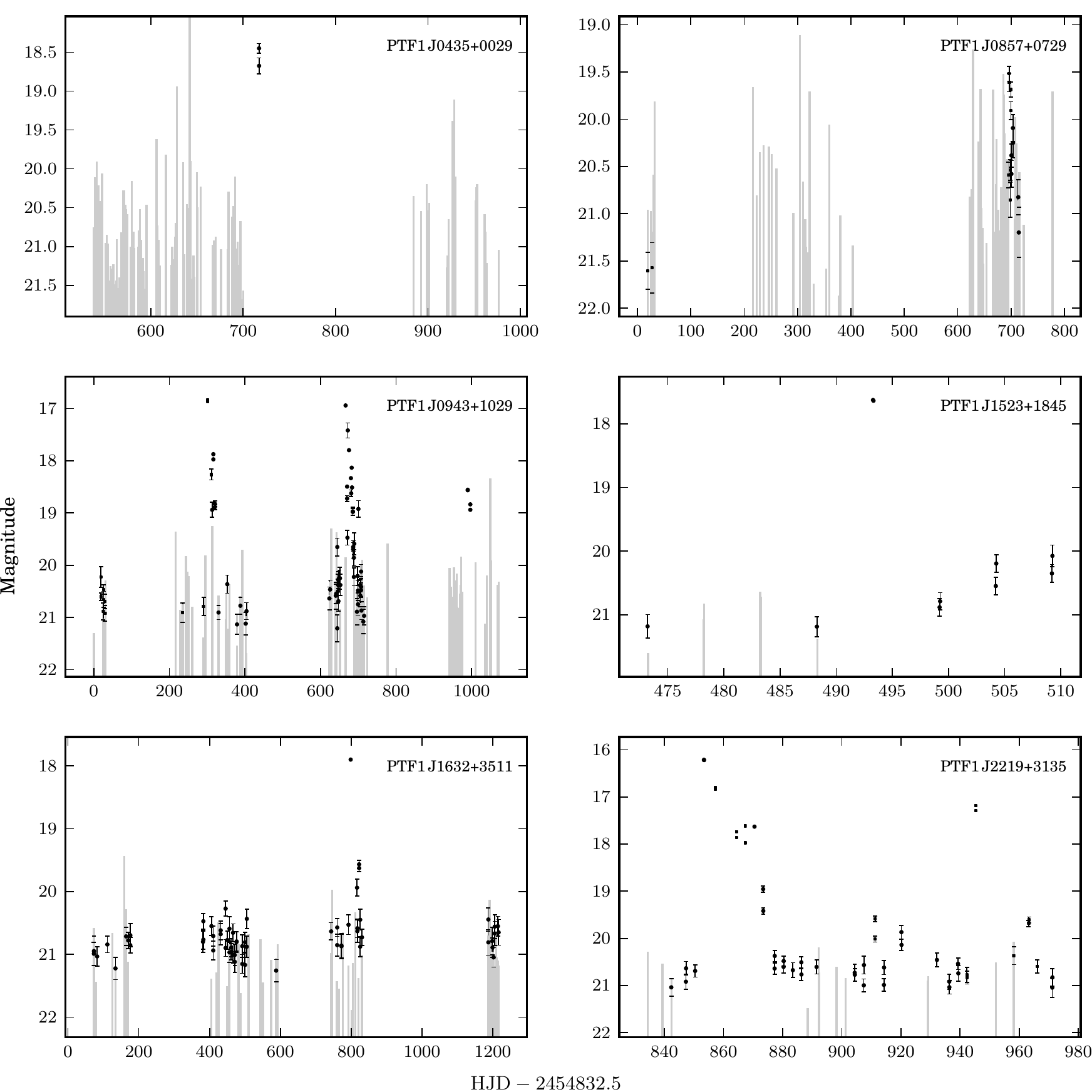}
\par\end{centering}
\caption{Light curves of all systems presented here. All data are from the PTF photometric pipeline.
Error bars are shown for those observations with errors $\gtrapp5\%$. Non-detections are indicated using gray lines
 -- the top of the line is the $3\sigma$ limiting magnitude as derived from the seeing, background, and CCD characteristics.
 Dates shown are relative to 01 Jan 2009. All data taken for each system by the PTF are shown except for PTF1J1523+1845, for which there
 were 12 prior non-detections to limiting magnitudes of $\simapp21$ over the 400 days before the data shown. We do not
 differentiate between $R$ and $g'$ data here, although the vast majority is $R$.}
\label{fig:lc}
\end{figure*}

We summarize the newly discovered AM CVn systems in Table \ref{tbl:systems}. Hereafter, we will
refer to all systems using the shorter PTF1JHHMM+DDMM convention as opposed to their full
coordinates\footnote{Non-transient sources in PTF are identified using the conventional IAU name format in the PTF1 catalog, a preliminary
version of the final PTF catalog. This is different from the PTF transient convention that identifies 
events by the year and a character sequence.}.
All observations are summarized in Table \ref{tbl:observations}. The PTF light curves for all discovered systems
are in Figure \ref{fig:lc}. We determined that these systems
are AM CVn systems based on the presence of helium, the lack of hydrogen, and the observed outbursts.

\begin{table*}
\begin{center}
\caption{PTF-Discovered AM CVn System Properties.}
\label{tbl:systems}
\begin{tabular}{@{}lcccccccc@{}}
\hline
\multicolumn{1}{c}{System} & Discovery & Period & Outburst & \multicolumn{4}{c}{Quiescent Magnitudes} \\
& Pipeline & (min) & Mag$^a$ & $u'$ & $g'$ & $r'$ & $i'$\\
\hline
PTF1\,J043517.73+002940.7$^{b}$ & Transient$^{c}$ & $34.31\pm1.94\,$ & $18.4\:(R)$ & $22.14\pm0.11$ & $22.28\pm0.04$ & $22.45\pm0.04$ & $22.60\pm0.10$\\
PTF1\,J085724.27+072946.7  & Transient$^{c}$ & \nodata & $19.5\:(R)$ & $21.68\pm0.02$ & $21.68\pm0.01$ & $21.74\pm0.02$ & $21.74\pm0.04$\\
PTF1\,J094329.59+102957.6  & Photometric & $30.17\pm0.65$ & $16.9\:(R)$ & $20.51\pm0.01$ & $20.71\pm0.01$ & $21.09\pm0.02$ & $21.17\pm0.05$\\
PTF1\,J152310.71+184558.2  & Transient$^{c}$ & \nodata & $17.6\:(R)$ & $23.28\pm0.12$ & $23.48\pm0.05$ & $23.34\pm0.06$ & $23.14\pm0.08$ \\
PTF1\,J163239.39+351107.3$^{b,d}$ & Transient$^{c}$ & \nodata & $17.9\:(g')$ & $22.74\pm0.14$ & $22.99\pm0.07$ & $22.98\pm0.06$ & \nodata \\
PTF1\,J221910.09+313523.1  & Photometric & \nodata & $16.2\:(g')$ & $20.50\pm0.03$ & $20.66\pm0.06$ & $20.90\pm0.03$ & $20.94\pm0.02$\\[1.5mm]
PTF1\,J071912.13+485834.0$^e$ & Transient & $26.77\pm0.02$ & $15.56\:(g')$ &\nodata & \nodata & \nodata & \nodata\\
\hline
\end{tabular}\end{center}
\begin{flushleft}
Quiescent magnitudes are from P200/LFC images and are not de-reddened.\\
$^a$ This is the brightest detection in PTF in either $R$ or $g'$ but is likely not the actual peak magnitude.\\
$^b$ Also identified as a CV candidate by the Catalina Real-Time Transient Survey \citep{2009ApJ...696..870D}: PTF1J0435+0029 = CSS090219:043518+00294; PTF1J1632+3511 = CSS110507:163239+351108.\\
$^c$ PTF transient names: PTF1J0435+0029 = PTF11avm; PTF1J0857+0729 = PTF11aab; PTF1J1523+1856 = PTF10noc; PTF1J1632+3511 = PTF11dkq.\\
$^d$ This is an AM CVn system candidate. See discussion in Section \ref{sec:16323511}.\\
$^e$ Originally published in \citet{2011ApJ...739...68L}. Included here for reference as a PTF-discovered AM CVn system.
\end{flushleft}
\end{table*}

\begin{table*}
\begin{center}
\caption{Details of Observations.}
\label{tbl:observations}
\begin{tabular}{cclccc}
\hline
System & Setup & \multicolumn{1}{c}{UT Date} & State & Gratings/Grisms & Exp. Time (s)\\
\hline
PTF1J0435+0029 & P200/DBSP & 2011 Mar 10 & Outburst & B: 600/4000, R: 158/7500 & 600\\
\ldots   & Keck-I/LRIS & 2011 Mar 12 & Outburst & B: 400/3400, R: 400/7500 & 800\\
\ldots  & Keck-I/LRIS & 2011 Mar 26 & Quiescence & B: 400/3400, R: 400/7500 & 1080\\
\ldots & Keck-I/LRIS & 2011 Oct 29$^a$ & Quiescence & B: 600/4000, R: 600/7500 & $180\,$s$ \times 81$\\
\ldots & P200/LFC & 2012 Nov 22 & Quiescence & Imaging ($u', g', r', i'$) & 480, except $u': 900$\\
PTF1J0857+0729  & KPNO\,4-m/RC & 2011 Feb 01$^a$ & Outburst & 316/4000 & 1800\\
\ldots  & NOT/ALFOSC & 2011 Feb 02 & Outburst& \#11 (200/5200) & 120\\
\ldots  & P200/LFC & 2011 Nov 30 & Quiescence & Imaging  ($g'$)  & $45\,$s$ \times 80$\\
\ldots & NOT/ALFOSC & 2012 Feb 28 & Quiescence & Imaging ($g'$)  & $60\,$s$ \times 110$\\
\ldots & P200/LFC & 2012 Nov 22 & Quiescence & Imaging ($u', g', r', i'$) & 300, except $u': 600$\\
PTF1J0943+1029  & Keck-I/LRIS & 2011 Oct 29 & Quiescence & B: 600/4000, R: 600/7500 & 1200\\
\ldots & Keck-I/LRIS & 2011 Dec 25$^a$  & Quiescence & B: 600/4000, R: 600/7500 & $180\,$s$ \times 80$\\
\ldots & Keck-I/LRIS & 2011 Dec 31$^a$  & Quiescence & B: 600/4000, R: 600/7500 & $180\,$s$ \times 25$\\
\ldots & P200/LFC & 2012 Nov 22 & Quiescence & Imaging ($u', g', r', i'$) & 300, except $u': 600$\\
PTF1J1523+1845  & Keck-I/LRIS & 2010 Jul 07& Outburst & B: 400/3400, R: 400/7500 & 200\\
\ldots  & Keck-I/LRIS & 2010 Jul 08$^a$  & Outburst & B: 400/3400, R: 400/7500 & 300\\
\ldots & P200/LFC & 2012 Jan 30 & Quiescence & Image($u', g', r', i'$) & 360, except $u': 540$\\
PTF1J1632+3511 & Keck-II/DEIMOS & 2011 Jul 05$^a$  & Unknown & 600ZD (600/7500) & 900\\
PTF1J2219+3135  & Keck-I/LRIS & 2011 Nov 25$^a$  & Quiescence & B: 400/3400, R: 400/7500 & 1200\\
\ldots & P200/LFC & 2012 Nov 22 & Quiescence & Imaging ($u', g', r', i'$) & 300, except $u': 600$\\
\hline
\end{tabular}\end{center}
\begin{flushleft}
$^a$ Indicates that this exposure (or the co-add of the exposures) is shown in this paper.\\
P200/DBSP: Palomar $200^{\prime\prime}$ telescope with the Double Spectrograph \citep{1982PASP...94..586O}.\\
Keck-I/LRIS: Keck-I 10-m telescope with the Low Resolution Imaging Spectrometer \citep{1995PASP..107..375O,1998SPIE.3355...81M}.\\
KPNO 4-m/RC: KPNO 4-m Mayall telescope with the RC Spectrograph.\\
NOT/ALFOSC: 2.5-m Nordic Optical Telescope with the Andalucia Faint Object Spectrograph and Camera.\\
P200/LFC: Palomar $200^{\prime\prime}$ telescope with the Large Format Camera.\\
Keck-II/DEIMOS: Keck-II 10-m telescope with the Deep Imaging Multi-Object Spectrograph \citep{2003SPIE.4841.1657F}.
\end{flushleft}
\end{table*}

\subsection{PTF1\,J043517.73+002940.7}

\label{sec:PTF1J0435+0029}

PTF1J0435+0029 was detected in outburst by the transient pipeline on
2011 Feb 16 and identified as a candidate of interest by the Galaxy Zoo Supernovae project \citep{2011MNRAS.412.1309S}.
Only two detections were made as the field was not observed again until the following season.
Follow-up classification spectra on 2011 Mar 10 and 2011 Mar 12 showed a mostly	
featureless continuum spectrum. However, a spectrum taken on 2011 Mar 26 showed helium emission
lines consistent with those of known AM CVn systems.

\begin{figure*}
\begin{centering}
\includegraphics{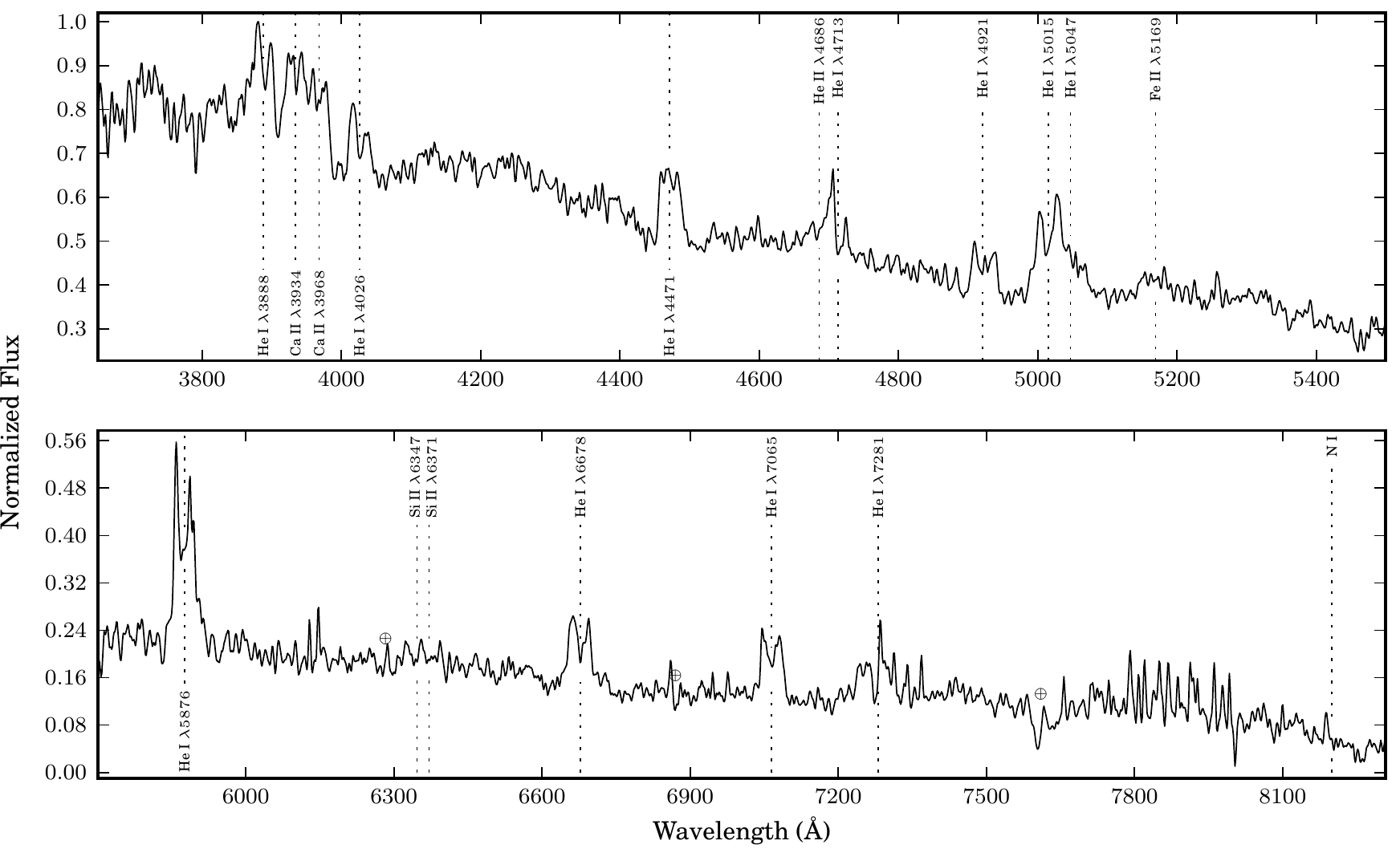}
\end{centering}
\caption{A spectrum of PTF1J0435+0029 obtained from the $\simapp4\,$h of co-added exposures of 2011 Oct 29.}
\label{fig:04350029:spec}
\end{figure*}

\begin{figure}
\begin{centering}
\includegraphics{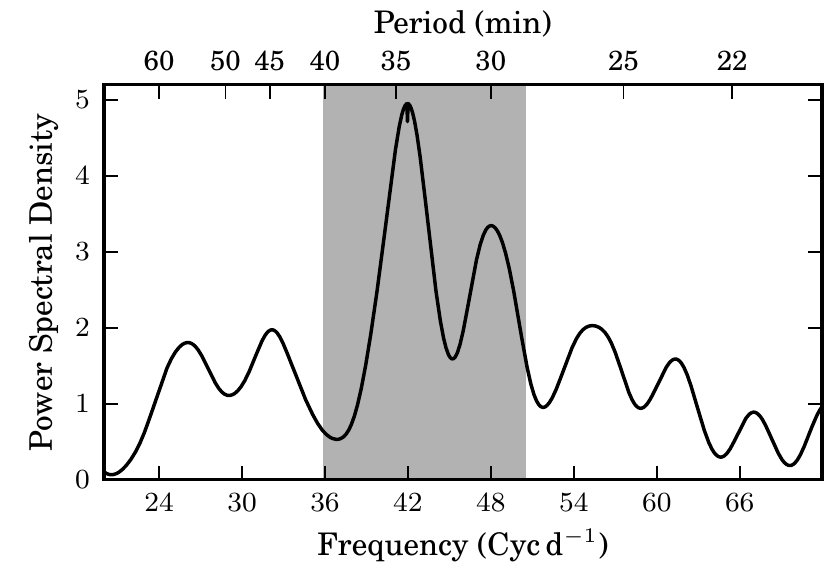}
\includegraphics{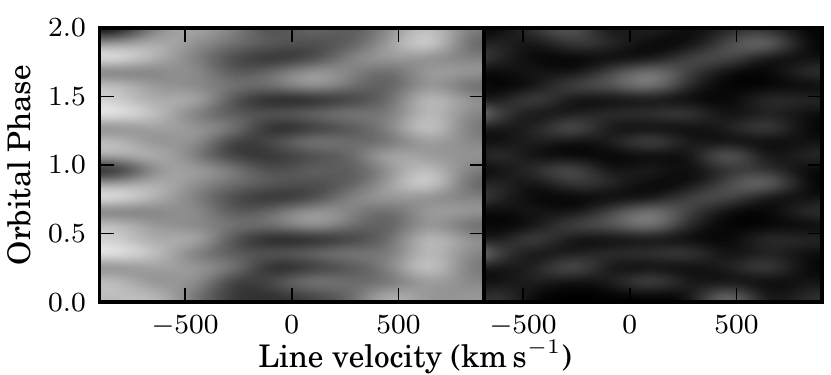}
\includegraphics{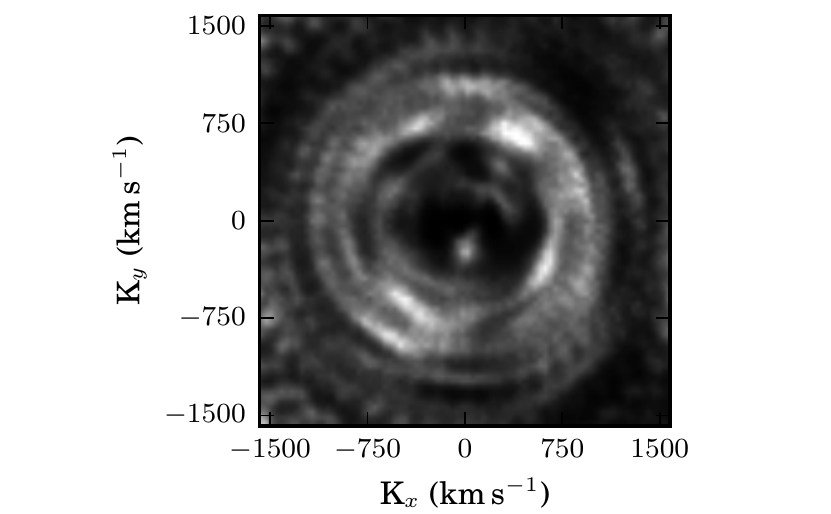}
\end{centering}
\caption{
{\bf Top:} Periodogram of the co-added \ion{He}{1} line flux ratios from 5.18\,h of PTF1J0435+0029 observations.
The calculation of these flux ratios is described in Section \ref{sec:peranal}. The \ion{He}{1}
lines at $\lambda\lambda 3888, 4026, 4471,\mathrm{~and~}5015$ were used to
calculate the flux ratios for this system. The $3\sigma$ confidence interval is shaded.
{\bf Middle:} The binned, trailed spectra of PTF1J0435+0029 using the co-added \ion{He}{1} lines at
$\lambda\lambda4026, 4471, 4713, 4921, 5015, 5875, 6678, \mathrm{~and~} 7065$ folded
at $34.31\,$min. This corresponds to the peak of the periodogram.
An arbitrary zero phase of $HJD=2455863.91002$ was used, coinciding
with the start of the observations. The version on the left retains the disc emission
while the version on the right removes the disc emission by subtracting the median of each column.
{\bf Bottom:} A Doppler tomogram of PTF1J0943+1029 constructed from the same emission lines as the S-wave,
plotted to highlight the peak believed to be the hotspot at $(K_x,K_y)\approx(380,655)\,\mathrm{km}~\mathrm{s}^{-1}$
(the upper right part of the image). An arbitrary zero phase of $HJD=2455863.91002$ was used, coinciding with the
start of the observations.}
\label{fig:04350029}
\end{figure}

On 2011 Oct 29, we obtained 5.18\,h of phase-resolved spectroscopy. We present the co-added spectrum in Figure \ref{fig:04350029:spec}.
We analysed this data as described
in Section \ref{sec:peranal} and present the periodogram of the flux ratios in Figure \ref{fig:04350029}. The periodogram
shows a peak at $34.31\,$min, but one that is broad due to the short baseline.
The uncertainty of period measurement was estimated using a simple Monte Carlo simulation. 
For each iteration of the simulation, we selected 105 exposures at random, allowing for repetition, and calculated the periodogram.
We estimate the error to be 1.94\,min, which is the standard deviation in the period estimates of 1000 such iterations. This is consistent
with the FWHM of the peak, which is 1.75\,min. The more complicated error estimate used for PTF1J0943+1029 (see Section \ref
{sec:PTF1J0943+1029}) could not be used here due to the lack of a strong signal.

\begin{figure}
\begin{centering}
\includegraphics{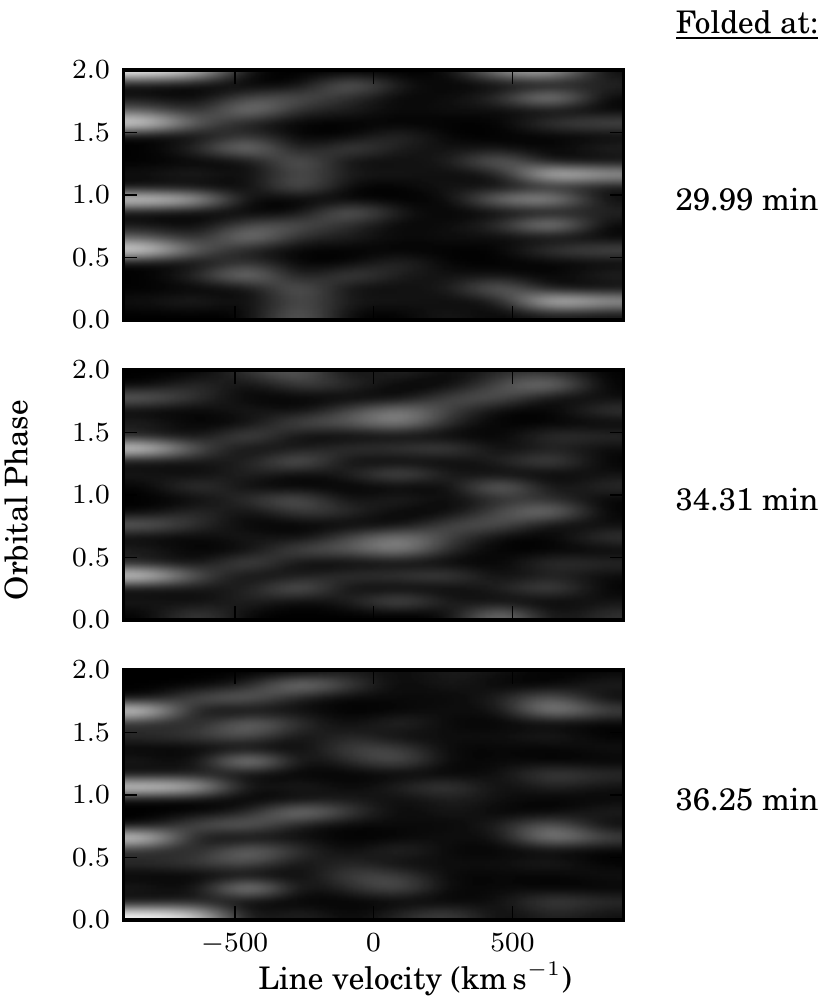}
\end{centering}
\caption{
A comparison of the binned, trailed spectra of PTF1J0435+0029 using the co-added \ion{He}{1} lines at
$\lambda\lambda4026, 4471, 4713, 4921, 5015, 5875, 6678, \mathrm{~and~} 7065$ folded,
from top to bottom, at $29.99\,$min, $34.31\,$min, and $36.25\,$min. These
correspond to the second most significant peak of the periodogram, the
proposed orbital period, and the period at the upper end of the error estimate, respectively.
An arbitrary zero phase of $HJD=2455863.91002$ was used for all three plots, coinciding
with the start of the observations. The median of each velocity bin was subtracted to remove
the contribution from the accretion disc. Only the plot at $34.31\,$min, the proposed
orbital period, shows a discernible S-wave; plots at other periods show noise and no signal.}
\label{fig:04350029:otherper}
\end{figure}

We trailed, binned, and folded the spectrum in an attempt to confirm the S-wave visually. This,
as well as a Doppler tomogram, is also presented in Figure \ref{fig:04350029}.
Figure \ref{fig:04350029:otherper} shows a comparison
of the S-wave at the stated orbital period to those generated with two other example periods:
$29.99\,$min and $P_{orb}+1.94\,$min. The former is the next highest peak on the periodogram.
The latter is one standard deviation away from the highest peak and would not be expected to show any signal
if the periodogram is valid. There is no S-wave present
at the alternate periods, despite the relatively high peak at $32.99\,$min in the periodogram.
Similar plots at other possible orbital periods likewise show no signs of an S-wave.

The proposed orbital period is in the same range as that of other known AM CVn systems with similar photometric behaviour.
\citet{2011ApJ...739...68L} and \citet{2012MNRAS.419.2836R} noted that systems with infrequent outbursts are associated with longer orbital periods.
The outburst on 2011 Feb 16 was the only observed outburst of PTF1J0435+0029 in PTF and there have been two recorded outbursts in the Catalina Real-Time Transient Survey \citep{2009ApJ...696..870D} -- one roughly coincidental with that observed by the PTF, and one $\simapp750$\,d prior.
We thus expect the system to have an orbital period between $\simapp27\,$min and $\simapp40\,$min (the longest observed
period for outbursting systems). The faintness of the S-wave
is not completely surprising since other systems \citep[e.g., SDSSJ0804+1616; ][]{2009MNRAS.394..367R} have also been observed
to have very weak S-waves at times.

We conclude our discussion of PTF1J0435+0029 with some remarks on the characteristics of its spectrum.
The spectrum of PTF1J0435+0029 is particularly notable in this set of AM CVn systems for the absence of
\ion{N}{1} (see Figure \ref{fig:04350029:spec} and Table \ref{tbl:metals}). This lack of \ion{N}{1}  likely points
to a different donor composition than the other systems presented here.
Specifically, the presence of \ion{N}{1} has been linked to a highly enriched CNO cycle in He-WD donors,
whereas the abundance decreases in He-star donors because of $\alpha$-capture on N \citep{2010MNRAS.401.1347N}.
Thus this system may have evolved from the He-star donor evolutionary track as opposed to the detached white dwarf binary track.

\subsection{PTF1\,J085724.27+072946.7}
\begin{figure*}
\begin{centering}
\includegraphics{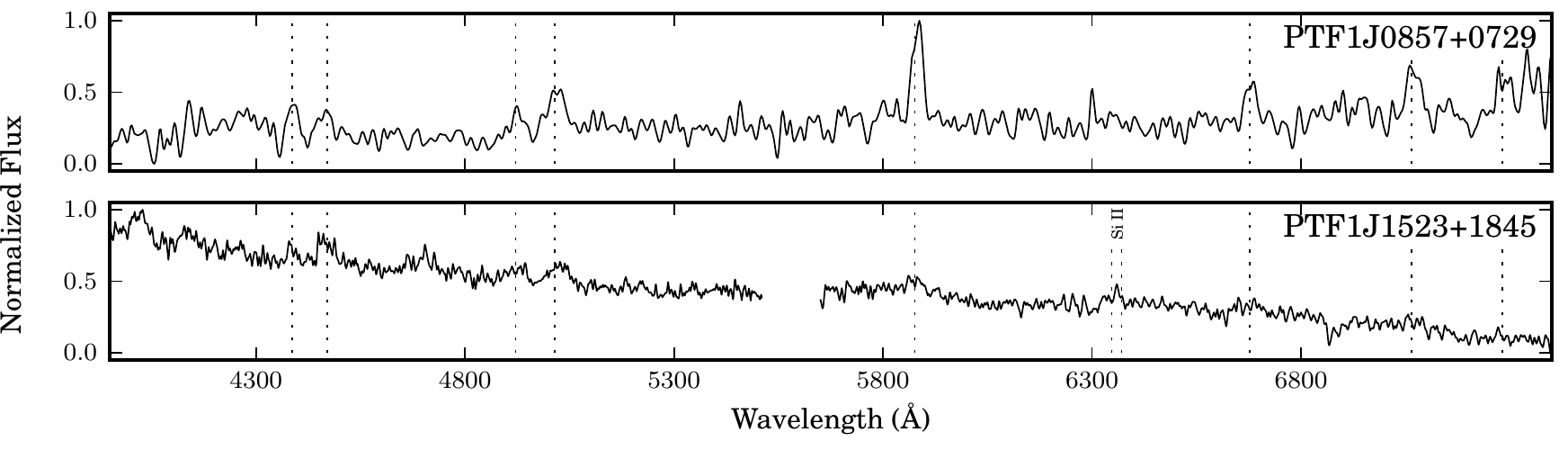}
\par\end{centering}
\caption{Spectra of PTF1J0857+0729 and PTF1J1523+1845 taken in outburst. Prominent \ion{He}{1} lines at $\lambda\lambda4387,
4471, 4921, 5015, 5875, 6678, 7065, $ and $7281$ are marked with dashed lines, as well as \ion{Si}{2} for
PTF1J1523+1845. The specific observations
used are marked in Table \ref{tbl:observations}. Spectra were Gaussian smoothed by 5 pixels. We particularly note the presence of
emission lines and no absorption lines in both systems, something that has not been observed previously in AM CVn
system outburst spectra.}
\label{fig:outburstspec}
\end{figure*}

PTF1J0857+0729 was discovered in outburst by the
transient pipeline on 2011 Jan 27. A classification spectrum taken on 2011 Feb 01 showed distinct helium emission lines,
which were confirmed with a second spectrum the following night (see Figure \ref{fig:outburstspec}).

\label{sec:PTF1J0857+0729}

In CVs, the presence of emission lines in outburst is indicative of a high inclination, and thus often eclipsing, system \citep[e.g., ][]{WARNER1995}.
Previously, all AM CVn systems spectroscopically observed in outburst showed absorption lines \citep[e.g., ][]{2007MNRAS.379..176R}.
We obtained a two-hour light curve at the P200 on 2011 Nov 30 (45\,s exposure time; 24\,s dead time)
and another two-hour light curve at the NOT on 2012 Feb 28 (60\,s exposure time; 5\,s dead time)
to search for eclipses. AM CVn systems are expected to have very short eclipses on the order of a minute \citep[e.g., ][]{2011MNRAS.410.1113C},
so short exposures times are necessary.

The individual photometric measurements had errors of $\simapp3\%$ and $\simapp5\%$, respectively. Periodograms constructed
from these light curves showed no significant period, although the light curves did show variability with an amplitude of $0.1$ -- $0.15\,$mag.
Lack of data precludes a definitive determination, but this is
consistent with the amplitude of periodic variability found in other quiescent AM CVn systems \citep[e.g., ][]{1997ApJ...480..383P}.
None of the photometric measurements were fainter than 0.1\,mag below the median magnitude and thus we conclude that no eclipses were detected.

We next consider whether eclipses may have been missed due to dead time between exposures. We assume that the eclipse duration is $60\,$s with
steep ingress and egress and a worst case scenario where the dead time is exactly in the middle of the eclipse.
Given the possible variability observed and the short observing times, we define an eclipse to be a measurement
at least $0.3\,$mag ($2$ -- $3\times$ the variability) below the median. We calculate that in this worst case scenario,
eclipses of $\gtrsimapp1.0\,$mag should have been visible in the P200 data and eclipses of $\gtrsimapp0.8\,$mag
should have been visible in the NOT data. Thus, any eclipses missed are relatively shallow.

\subsection{PTF1\,J094329.59+102957.6}

\begin{figure*}
\begin{centering}
\includegraphics{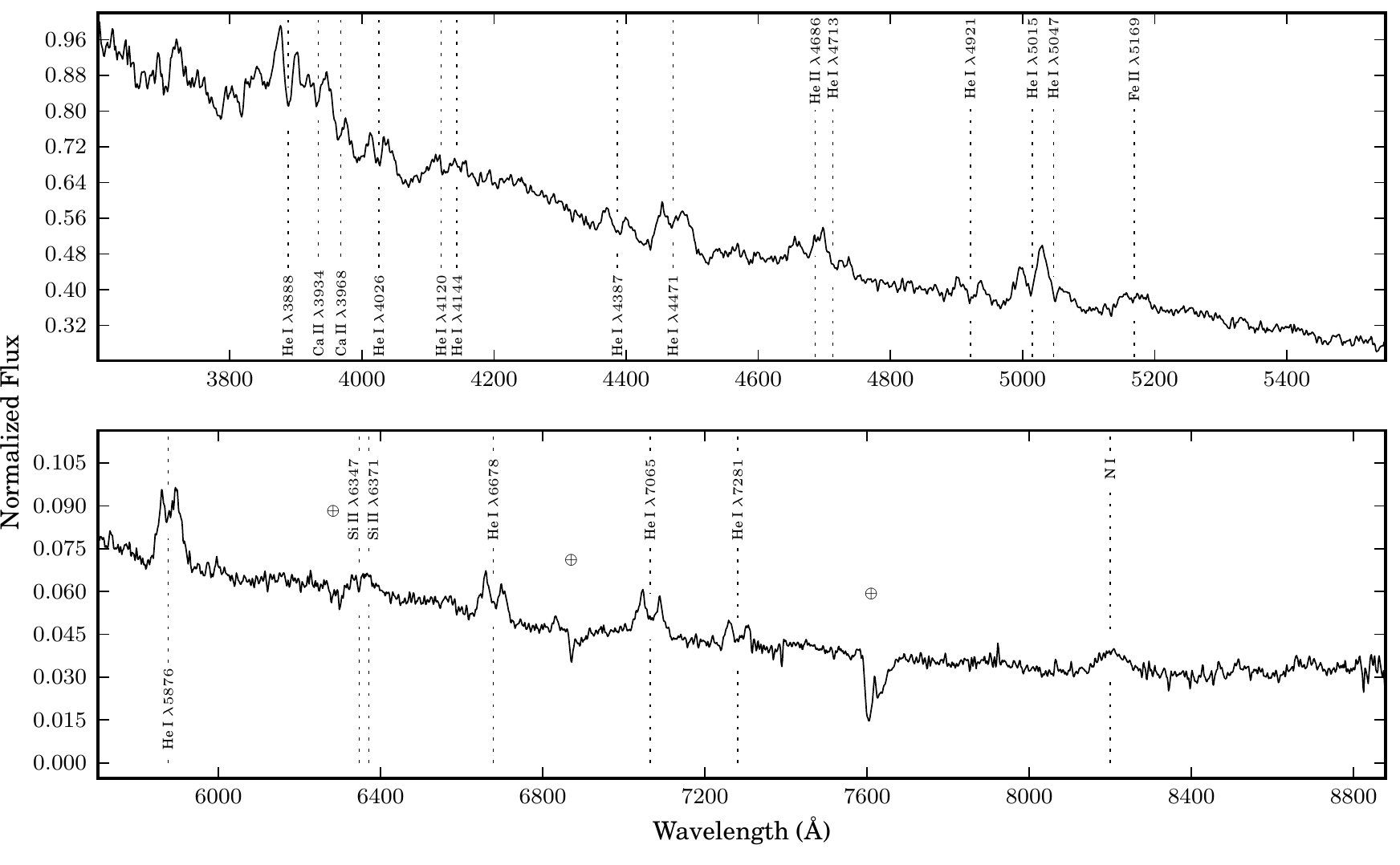}
\end{centering}
\caption{A spectrum of PTF1J0943+1029 in quiescence. The blue (top) spectrum is a co-add of the $\simapp4\,$h observed on 2011 Dec 25.
The red (bottom) spectrum is a co-add of the $\simapp1.5\,$h observed on 2011 Dec 31.}
\label{fig:09431029:spec}
\end{figure*}

PTF1J0943+1029 was discovered as part of the photometric database search (see light curve in Figure \ref{fig:lc}).
An initial classification spectrum showed strong helium emission lines. We obtained a total of 5.45\,h of phase-resolved spectroscopy 
of PTF1J0943+1029 on 2011 Dec 25 and 2011 Dec 31 using Keck-I/LRIS.
We present the co-added spectrum in Figure \ref{fig:09431029:spec}.

\begin{figure}
\begin{centering}
\includegraphics{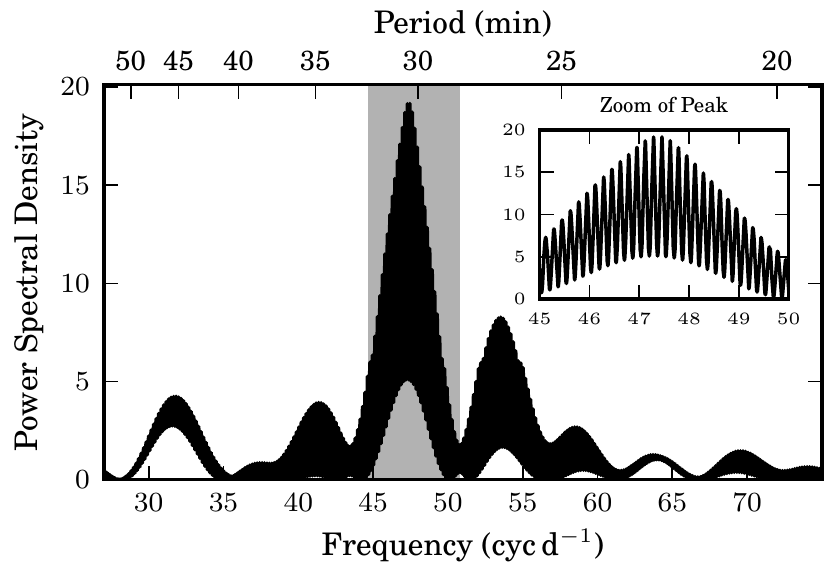}
\includegraphics{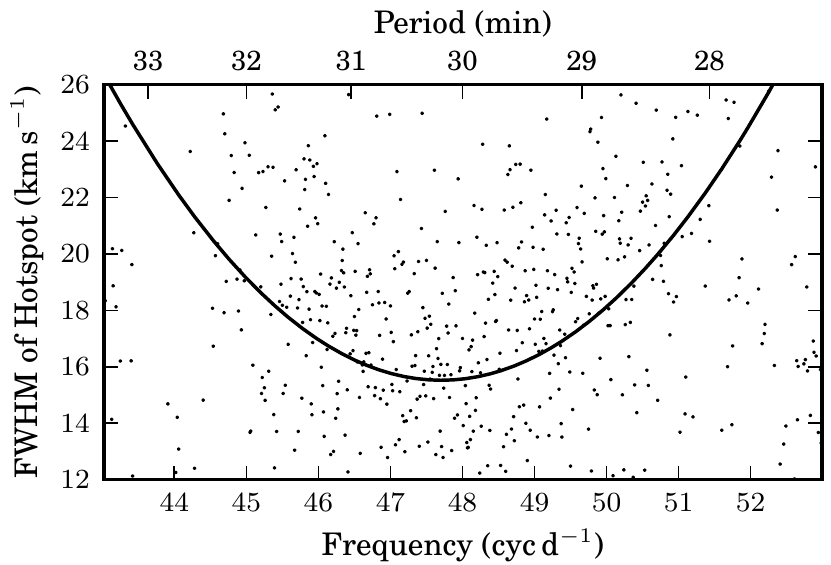}
\end{centering}
\caption{{\bf Top:} Periodogram of the flux ratios of PTF1J0943+1029. There are no strong peaks outside of the frequency range shown here.
The peak is at $30.35\,$min. The \ion{He}{1}
lines at $\lambda\lambda 3888, 4026, 4471,\mathrm{~and~}5015$ were used to
calculate the flux ratios for this system. The shaded region represents the $3\sigma$ confidence interval around the proposed orbital period.
{\bf Bottom:} The FWHMs of the hotspots in 1,000 Doppler maps calculated at a range of periods. The solid line is the best fit
of a quadratic equation. Its minimum of 30.17\,min is within one sigma of the peak of the periodogram.}
\label{fig:09431029:pr}
\end{figure}

\begin{figure}
\begin{centering}
\includegraphics{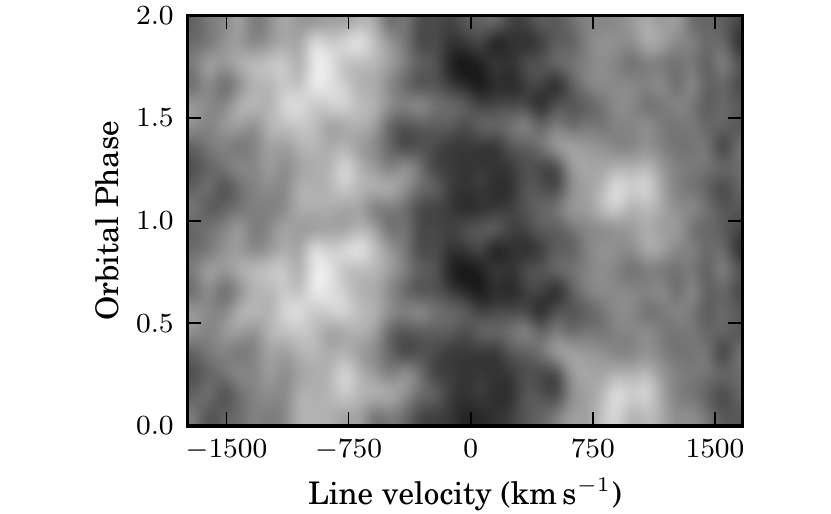}
\includegraphics{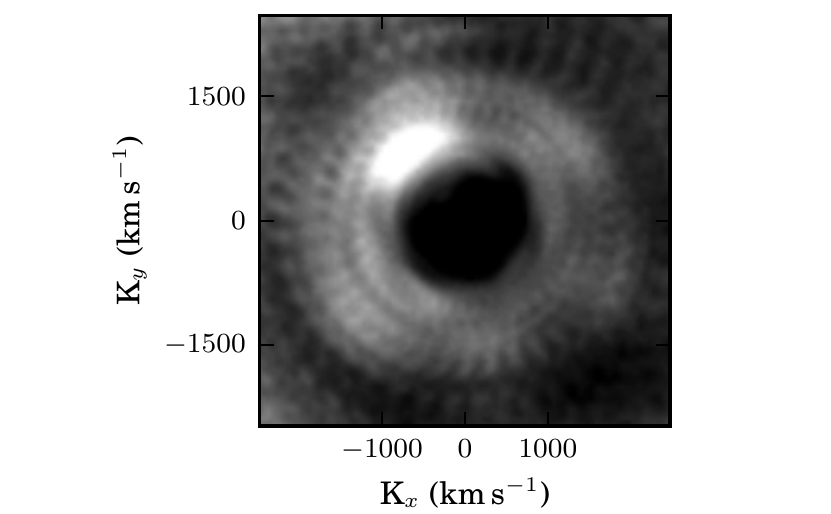}
\end{centering}
\caption{{\bf Top:} The S-wave used to visually confirm the period of PTF1J0943+1029. The \ion{He}{1} lines at
$\lambda\lambda3888, 4026, 4471, 4713, 4921, \mathrm{~and~} 5015$ and the \ion{He}{2} line at $\lambda4686$
were folded at a period of $30.35$ min
to generate the image. The S-wave was phase-binned into 10 bins with a zero phase
of $\mathrm{HJD}= 2455920.962$.  {\bf Bottom:} Doppler tomogram of PTF1J0943+1029 using
the \ion{He}{1} $\lambda\lambda 4027, 4471, 4713, 4921, \mathrm{~and~}5015$ and \ion{He}{2} $\lambda4685$.
The zero phase here is the same as for the S-wave.}
\label{fig:09431029:swave}
\end{figure}

This data were analysed for the orbital period. The peak of the periodogram is at $30.35\,$min (Figure \ref{fig:09431029:pr}), and an
S-wave and Doppler tomogram generated at this period show a strong signal (Figure \ref{fig:09431029:swave}). The number of strong
aliases adjacent to the peak frequency make a good error estimate crucial.

To obtain an estimate of the error, we exploited the properties of the Doppler tomograms. As discussed in Section \ref{sec:peranal},
the hotspot should be concentrated in a single spot on the Doppler tomogram. Hence, the correct orbital period should correspond to the
Doppler tomogram with the sharpest hotspot. We define this to be the hotspot with the smallest FWHM, using a two-dimensional Gaussian model.
We note that this method could not be used reliably for PTF1J0435+0029 due to the much worse signal-to-noise in that data.

We calculated 1,000 Doppler tomograms for periods between 27.2\,min and 33.5\,min in $\simapp0.1\,$min steps (43 to 53\,cyc/d in 0.1\,cyc/d steps)
and measured the FWHM of the hotspot in each using a two-dimensional Gaussian fit. To estimate the error, for each of 1,000 iterations, we drew 100 FWHM measurements from the range of 46 to 50 cyc/d and fit a parabola to the measurements. This limited range was required to eliminate inaccurate measurements due to multi-peak ``spots'' outside of this range that gave inaccurate, small FWHM measurements.

This simulation found a median period of $30.17\pm0.65\,$min, within one sigma of the peak of the periodogram. We plot all FWHM's and the best
fit using all points in Figure \ref{fig:09431029:pr}. The error estimate of this simulation is consistent with a visual inspection of S-waves and Doppler 
tomograms around the peak of the periodogram. Given the error and the visual representations of the orbit in Figure \ref{fig:09431029:swave}, we conclude that the median period from this simulation is the orbital period of the system.

\label{sec:PTF1J0943+1029}

 \subsection{PTF1\,J152310.71+184558.2}
 
 PTF1J1523+1845 was discovered in outburst on 2010 Jul 7 as part of the transient search. Spectra taken on 2010 Jul 7 and 2010 Jul 8
 showed \ion{He}{1} emission lines (see Figure \ref{fig:outburstspec}). This makes it the second known outbursting AM CVn system to exhibit emission lines in
 outburst. Extensive follow-up on this system was not performed, due to its faint nature ($g'>23$).
 
\subsection{PTF1\,J163239.39+351107.3}
\label{sec:16323511}

\begin{figure}
\begin{centering}
\includegraphics{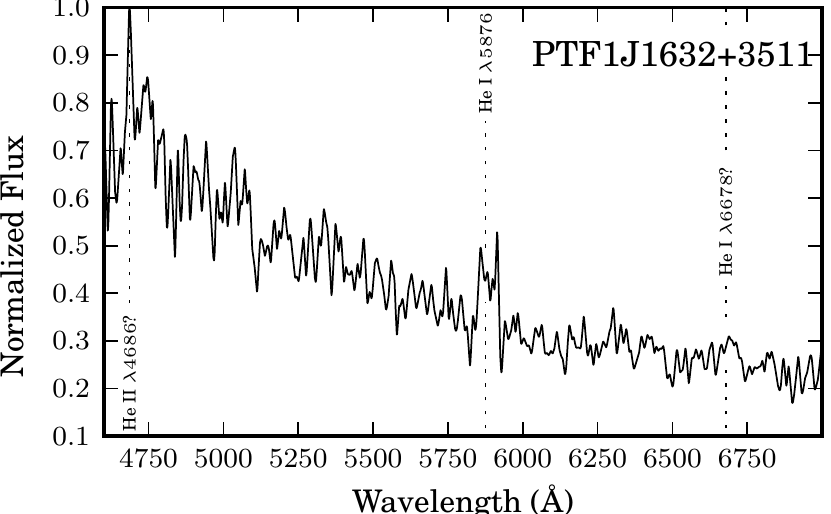}
\par\end{centering}
\caption{Classification spectrum of PTF1J1632+3511 taken on 2011 Jul 5. The system is assumed to be either in or near quiescence.
The closest PTF photometric measurement on 2011 Jul 11 has no detection to a limiting magnitude of 20.67.
The poor quality of the spectrum precludes us from classifying this as an AM CVn system. However, after Gaussian smoothing by 10 pixels
we can identify \ion{He}{1} $\lambda5875$ emission, possible \ion{He}{1} $\lambda6678$ and \ion{He}{2} $\lambda4686$
emission, and no evidence of H.}
\label{fig:16323511spec}
\end{figure}

PTF1J1632+3511 was discovered in outburst on 2011 May 11 and a spectrum was obtained on 2011 Jul 5.
This source is extremely faint ($g'\approx23$) and is located adjacent to a significantly brighter galaxy. We present
the spectrum in Figure \ref{fig:16323511spec}.
The spectrum shows \ion{He}{1} $\lambda 5875$ emission, possible \ion{He}{1} $\lambda6678$ and \ion{He}{2} $\lambda4686$ emission,
and no trace of H lines. We exclude it as a possible supernova from the \ion{He}{1} lines at $z=0$.
Based on the evidence, we conclude that PTF1J1632+3511 is a likely AM CVn system, but a higher signal-to-noise spectrum
is required before this can be established with certainty.

 \subsection{PTF1\,J221910.09+313523.1}
 
\begin{figure*}
\begin{centering}
\includegraphics{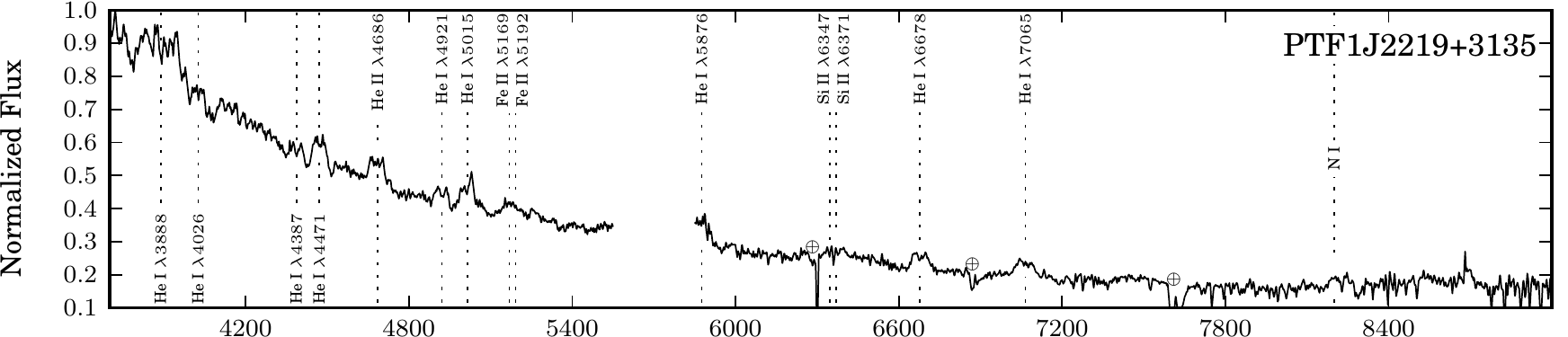}
\par\end{centering}
\caption{Classification spectrum of PTF1J2219+3135 taken in quiescence. The prominent He and metal lines are marked (see Section \ref{sec:specfeat}). }
\label{fig:22193135spec}
\end{figure*}

PTF1J2219+3135 was discovered as part of the photometric database search (see Figure \ref{fig:lc} for the light curve and
Figure \ref{fig:22193135spec} for the identification spectrum).
The relatively low quality identification spectrum has a particularly interesting set of lines. 
Specifically, several lines redward of \ion{Fe}{2} $\lambda 5169$ at $\lambda5276$ and $\lambda5317$ have only been observed
in V406 Hya \citep{2006MNRAS.365.1109R} ($P_\mathrm{orb}=33.8\,$min) and SDSSJ0804+1616 \citep{2009MNRAS.394..367R} ($P_\mathrm{orb} = 44.5\,$min). These may also be \ion{Fe}{2}, but
such lines are unusual.
Note that unique identification of weak features is made more difficult due to the large width of the lines due to
Doppler broadening of the rotating accretion disc.

\section{Discussion}
\label{sec:Discussion}

\subsection{Spectral Features}\label{sec:specfeat}

We present equivalent widths for lines in each of the five systems presented here in Table \ref{tbl:metals}.
In this section, we compare the features of the systems with relatively high signal-to-noise: PTF1J0435+0029, PTF1J0943+1029, and PTF1J2219+3135.
As expected, the spectra of the systems in this sample show signs of having shorter orbital periods than those
detected in the SDSS. Particularly, we note the lack or relative weakness of the absorption wings surrounding the He
emission lines. These absorption wings are from the WD primary and are expected to be less visible in shorter period systems
if the mass transfer rate monotonously decreases with increasing orbital period, since this will create a relatively luminous accretion
disc with respect to the luminosity of the accretor. We also see the presence of \ion{Ca}{2} H \& K in all three systems, which has been seen in
shorter period systems such as V406 Hya \citep{2006MNRAS.365.1109R}, but is less prevalent in longer period systems
\citetext{those with $P_{orb}>50\,$min; \citealp{2005MNRAS.361..487R}; Kupfer et al. in prep}.

\begin{table*}
\begin{center}
\caption{Equivalent Widths of Prominent Lines.}
\label{tbl:metals}
\begin{tabular}{cccccc}
\hline
Line & PTF1J0435+0029 & PTF1J0857+0729 & PTF1J0943+1029 & PTF1J1523+1845 & PTF1J2219+3135\\
\hline
\ion{Ca}{2} 3933/3968    &    blended with He   & X & blended with He  & X & blended with He  \\
\ion{He}{1}  4026        &  -4.6 $\pm$ 0.7  & X &   X  & -3.7 $\pm$ 0.9 &  -5.1 $\pm$ 0.8 \\
\ion{He}{1}  4388        &  X              & \nodata$^b$ &   -2.3 $\pm$ 0.3  & -1.2 $\pm$ 0.5 &  -1.7 $\pm$ 0.2 \\
\ion{He}{1}  4471        &  -10.0 $\pm$ 0.7 & -13 $\pm$ 5.1 &  -10.4 $\pm$ 0.3  & -5.7 $\pm$ 0.7 &  -6.6 $\pm$ 0.2  \\
\ion{He}{2} 4685/4713   &   -6.4 $\pm$ 0.8 & X &  -9.6 $\pm$ 0.4 & -5.7 $\pm$ 0.9 &   -3.5 $\pm$ 0.3 \\
\ion{He}{1} 4921        &  -7.9  $\pm$ 0.7 & -10.6 $\pm$ 5.4 &   -1.6 $\pm$ 0.3  & -2.5 $\pm$ 0.7 &   -4.8 $\pm$ 0.2 \\
\ion{He}{1}  5015/5047   &  -10.0 $\pm$ 0.6 & -34.9 $\pm$ 5.9  & -17.1 $\pm$ 0.4  & -5.5 $\pm$ 0.8  &  -9.1 $\pm$ 0.2  \\
\ion{He}{1}  5875        &  -49.8 $\pm$ 2.0 &  -62.9 $\pm$ 5.6 & -21.8 $\pm$ 0.6  & -7.4 $\pm$ 0.7 &   X  \\
\ion{He}{1}  6678        &  -33.1 $\pm$ 2.0 & -19.6 $\pm$ 5.6  &  -14.1 $\pm$ 0.6  & -4.2 $\pm$ 0.8 &  -10.7 $\pm$ 0.2  \\
\ion{He}{1} 7065        &  -29.8 $\pm$ 1.9 & -26.6 $\pm$ 5.1 &  -15.0 $\pm$ 0.7  & \nodata$^b$ &  -15.1 $\pm$ 0.3 \\
\ion{He}{1}  7281        &  -28.4 $\pm$ 2.4 & X & -9.3 $\pm$ 0.6\  &  X  & \nodata$^a$  \\
\ion{Fe}{2} 5169        &          -0.8 $\pm$ 0.7   & X &   -4.8 $\pm$ 0.3  & \nodata$^b$ &  -9.8 $\pm$ 0.2  \\
\ion{Si}{2} 6347/6371   &          \nodata$^b$        & X  & -6.5 $\pm$ 0.7  & -3.8 $\pm$ 0.6  &  -9.5 $\pm$ 0.3   \\
\ion{N}{1} 8184/8188/8200  &     X     &   \nodata$^c$ &   -7.8 $\pm$ 0.7   &  \nodata$^a$ & -3.6 $\pm$ 0.9\\
\hline
\end{tabular}\end{center}
\begin{flushleft}
Lines marked with an X indicate that this line is not detectable above the noise level of the spectrum obtained.\\
$^a$ Line present but contaminated with atmosphere.\\
$^b$ Line present but insufficient SNR to measure.\\
$^c$ Spectrum does not extend to this wavelength.
\end{flushleft}
\end{table*}

Of particular interest for the general study of AM CVn systems is the identification of the donor and
therefore the evolutionary history of these systems. \citet{1991ApJ...366..535M} predicted the \ion{Si}{2}
emission at $\lambda6346$ and $\lambda6371$ and \ion{Fe}{2} emission at $\lambda5169$
to be the strongest lines in helium dominated optically thin accretion discs. They
will be important to determine the initial metallicity in follow-up work as iron and silicon are not supposed to be
affected by nuclear synthesis processes in AM CVn systems. Those features are well observed in other outbursting AM CVn 
systems, including V406 Hya \citep{2006MNRAS.365.1109R} and CP Eri \citep{2001ApJ...558L.123G}.
Here, we find the presence of \ion{Si}{2} and \ion{Fe}{2} in all three systems. However, as noted earlier,
\ion{N}{1} is uniquely missing in PTF1J0435+0029, indicating the donor is more likely to have evolved from a He-star.
PTF1J0943+1029 and PTF1J2219+3135, on the other hand, are more likely to have a He WD as the donor star.

\subsection{Comparison of System Colours}

The release of the SDSS data revolutionized the study of AM CVn systems. Initially, seven
systems were discovered via a search for He emission lines in the SDSS database \citep{2005MNRAS.361..487R,2005AJ....130.2230A, 2008AJ....135.2108A}.
Subsequently, a spectroscopic survey of a colour-selected sample from SDSS has been carried out and six more systems were found
\citep{2009MNRAS.394..367R,2010ApJ...708..456R,CARTERAMCVNS}.
However, both of these studies have used colours to select systems for spectroscopic observation. In the case of
the general SDSS survey, various criteria are used to select follow-up targets while the dedicated AM CVn search
used a colour cut based on the colours of known AM CVn systems.

The sample here, together with PTFJ0719+4858 presented by \citet{2011ApJ...739...68L}, is unique as the first sample of
AM CVn systems discovered systematically by their large-amplitude photometric variability as opposed to their spectral characteristics
(and equivalently colours). It thus allows us to consider the completeness of the colour-selection criteria developed in \citet{2009MNRAS.394..367R}.

We note that these criteria target only quiescent systems since the spectra and colours of AM CVn systems do change during outburst.
Given that the SDSS survey is expected to be most sensitive to longer-period systems and that these are believed to make up
the vast majority of the population, this is a valid assumption. However, although our colors of PTF-discovered systems
are from the quiescent states, our better sensitivity to shorter period systems may result in slightly different colors.

\begin{figure*}
\begin{centering}
\includegraphics{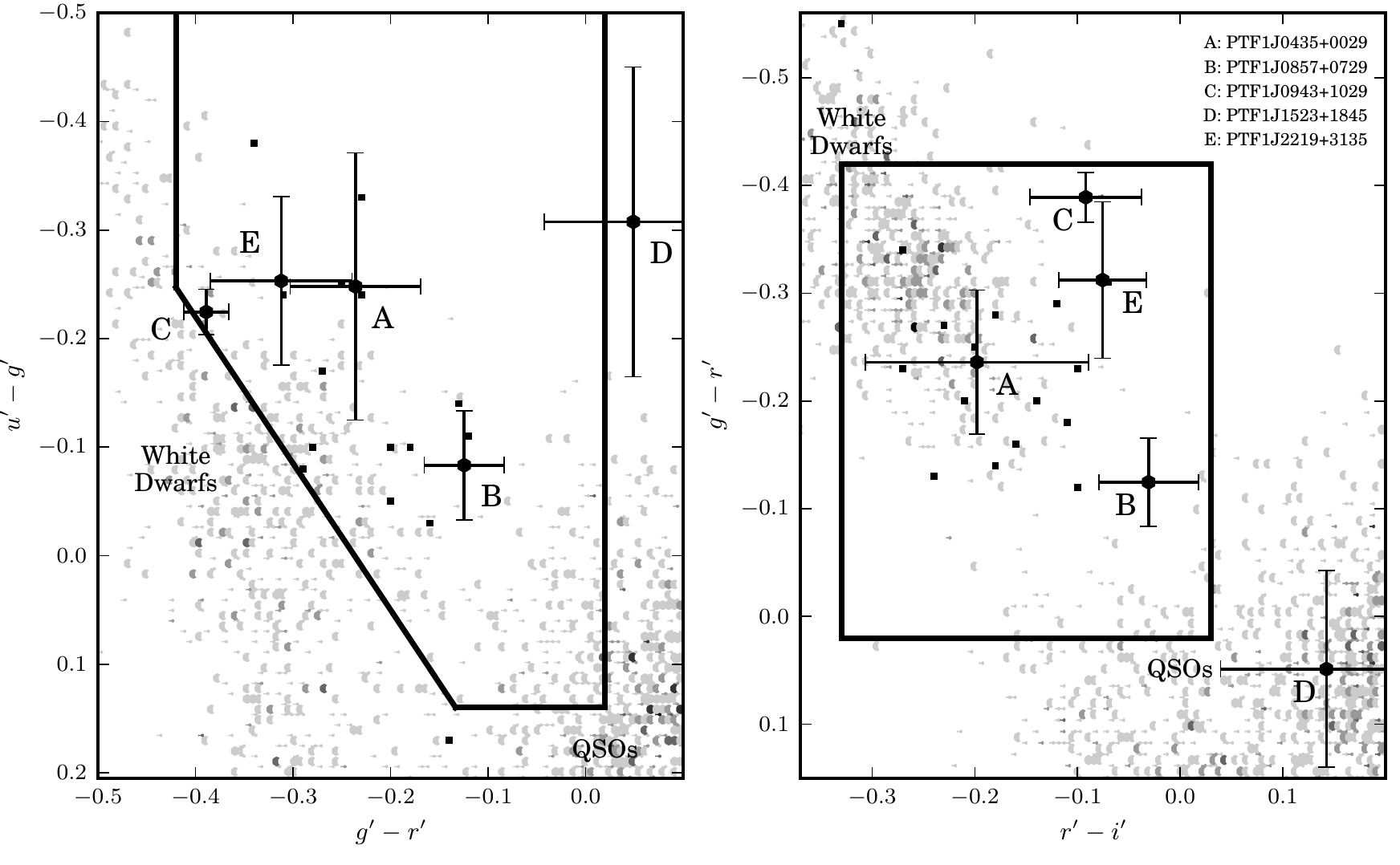}
\end{centering}
\caption{The colour-colour diagrams for AM CVn systems, including background SDSS-detected systems. AM CVn
system are marked as black squares. Those labelled with a letter and with error bars are from this paper; those
without are previously known systems with SDSS
colours. Shaded areas indicate the density of the background systems -- darker areas indicate
a denser area in colour-colour space. The lines delineate the colour cut from  \citet{2009MNRAS.394..367R}.
We have corrected for extinction in the same way as \citet{2009MNRAS.394..367R}, but note that the distances to AM CVn
systems are relatively small \citep{2007ApJ...666.1174R} and thus the reddening is likely
overestimated \citep{1998ApJ...500..525S,2009MNRAS.394..367R}.
The PTF-discovered systems tend to lie close to the colour cut, but their exact position is uncertain due to large error bars.}
\label{fig:colours}
\end{figure*}

The quiescent magnitudes in each filter were presented in Table \ref{tbl:systems} and we plot the colours of our sources,
together with known AM CVn systems and background sources in Figure \ref{fig:colours}.
The colours of the PTF-discovered systems are from the quiescent state; those of previously known systems
are determined from the SDSS photometry and may or may not be in quiescence. However, given that these are all longer-period systems,
it is highly likely that almost all are in quiescence. We do not include PTF1J1635+3511 in this plot since it is only a candidate.
Additionally, it is likely contaminated with the nearby galaxy, especially at the redder wavelengths.

We find that the PTF-discovered systems generally fall within the colour cut. One system, PTF1J1523+1845, 
does show colours outside of the colour cut, but, given its faintness, additional measurements should be made to verify this.
Given these results, we urge caution in narrowing the colour cut, as suggested by \citet{CARTERAMCVNS}, to avoid
missing redder AM CVn systems. However, both additional systems and further measurements are necessary before
concluding whether outburst-identification is sensitive to a population substantially different than the SDSS colour-selected sample.

\section{Conclusions}

We present five new AM CVn systems and one new AM CVn candidate, which were identified by their characteristic
\ion{He}{1} and \ion{He}{2} emission lines and lack of H. We further present spectroscopic
measurements of three of these systems, finding their orbital periods to be consistent with
other AM CVn systems.
Two of the systems presented here have the unique
characteristic of strong emission lines while in outburst, often seen in eclipsing cataclysmic
variable systems. We tested whether one of these systems is eclipsing but did not find evidence to support this hypothesis.
Finally, we compared the spectroscopic and photometric features of these systems to other known
AM CVn systems.

\section*{Acknowledgments}

We thank Sagi Ben-Ami, Yi Cao, Brad Cenko, Avishay Gal-Yam, Assaf Horesh, Mansi Kasliwal, Thomas Matheson,
Kunal Mooley, and Robert Quimby for help in obtaining observations and reducing data.
We thank Kevin Rykoski and Carolyn Heffner at the Palomar Observatory
for developing the fast cadence mode on the LFC instrument.
Part of this work was performed by TAP while
at the Aspen Center for Physics, which is supported by
NSF Grant \#1066293. PJG thanks the California Institute of Technology for
its hospitality during his sabbatical stay.

Observations obtained with the Samuel Oschin Telescope at the Palomar
Observatory as part of the Palomar Transient Factory project, a scientific
collaboration between the California Institute of Technology, Columbia
University, Las Cumbres Observatory, the Lawrence Berkeley National
Laboratory, the National Energy Research Scientific Computing Center,
the University of Oxford, and the Weizmann Institute of Science. Some
of the data presented herein were obtained at the W.M. Keck Observatory,
which is operated as a scientific partnership among the California
Institute of Technology, the University of California and the National
Aeronautics and Space Administration. The Observatory was made possible
by the generous financial support of the W.M. Keck Foundation. 
The authors wish to recognize and acknowledge the very significant cultural role and reverence that the summit of Mauna Kea has always had within the indigenous Hawaiian community.  We are most fortunate to have the opportunity to conduct observations from this mountain. 
Based in part on observations made with the Nordic Optical Telescope, operated
on the island of La Palma jointly by Denmark, Finland, Iceland,
Norway, and Sweden, in the Spanish Observatorio del Roque de los
Muchachos of the Instituto de Astrofisica de Canarias.
The data presented here were obtained in part with ALFOSC,
which is provided by the Instituto de Astrofisica de Andalucia (IAA)
under a joint agreement with the University of Copenhagen and NOTSA.
This research has made use of NASA's Astrophysics Data System.

\bibliographystyle{mn2e}
\bibliography{amcvns}

\end{document}